\documentclass[paper]{geophysics}
\pdfoutput=1
\usepackage{amsmath}
\usepackage{amssymb}
\usepackage{amsfonts}
\usepackage{color}
\usepackage{array}
\usepackage{seg}

\begin{document}

\renewcommand{\figdir}{.} 

\title{Robust joint full-waveform inversion of time-lapse seismic data sets with total-variation regularization}
\author{Musa Maharramov (maharram@stanford.edu)\\ Biondo Biondi (biondo@sep.stanford.edu)}
\righthead{Joint 4DFWI}
\lefthead{Maharramov and Biondi}
\footer{SEP--155}
\maketitle

\begin{abstract}
        We present a technique for reconstructing subsurface velocity model changes from time-lapse seismic survey data using full-waveform inversion (FWI). The technique is based on simultaneously inverting multiple survey vintages, with model difference regularization using the total variation (TV) seminorm. We compare the new TV-regularized time-lapse FWI with the $L_2$-regularized joint inversion proposed in our earlier work, using synthetic data sets that exhibit survey repeatability issues.  The results demonstrate clear advantages of the proposed TV-regularized joint inversion over alternatives methods for recovering production-induced model changes that are due to both fluid substitution and geomechanical effects. 
\end{abstract}

\section{Introduction}
Effective reservoir monitoring depends on successful tracking of production\--induced fluid movement in the reservoir and overburden, using input from seismic imaging, geomechanics, geology and reservoir simulation \cite[]{Biondi96}. To achieve this, most traditional methods rely on the conversion of picked time shifts and reflectivity differences between migrated images into reflector movement and impedance changes. Though effective in practical applications, this approach requires a significant amount of expert interpretation and relies on quality control in the conversion process. Wave\--equ\-ation image-difference tomography has been proposed as a more automatic alternative method to recover velocity changes \cite[]{UweMe}; it allows localized target-oriented inversion of model perturbations \cite[]{MeUwe}. An alternative approach is based on using the high-resolution power of the full-waveform inversion \cite[]{SirgueFirstBreak} to reconstruct pro\-duc\-tion\--induced changes from wide-offset seismic acquisitions, and is the subject of this paper.

Time-lapse full-waveform inversion \cite[]{Watanabe,DenliHuang,Routh} is a promising technique for time-lapse seismic imaging where produc\-tion-induced subsurface model changes are within the resolution of FWI. However, as with alternative time-lapse techniques, time-lapse FWI is sensitive to repeatability issues \cite[]{Amir}. Non-repeatable acquisition geometries (e.g., slightly shifted source and receiver positions), acquisition gaps (e.g., due to new obstacles), different source signatures and measurement noise---all contribute to differences in the data from different survey vintages. Differences in the input data sets due to repeatability issues may easily mask valuable production\--induced changes. However, even with noise-free synthetic data without any acquisition repeatability issues, numerical artifacts may contaminate the inverted difference of monitor and baseline when practical limitations are imposed on solver iteration count. \cite{MusaSEP150} devised a time-lapse FWI that minimized model differences outside of areas affected by production by jointly inverting for multiple models, and imposing a regularization condition on the model difference. The joint inversion can be performed simultaneously for multiple model vintages or using an empirical technique of ``cross-updating'' \cite[]{MusaSEP150,MusaSEP152}. \cite{MusaSEP152,MusaAGUSEG14} applied these methods to noisy synthetic data and compared the results to alternative methods, demonstrating advantages of both target-oriented simultaneous inversion and cross-updating over alternative methods. Production-induced changes that are due to fluid substitution effects are spatially constrained to areas affected by fluid transport, while those due to geomechanical effects---e.g., stretching of the overburden---may result in smoothly varying velocity model differences. In either case, regularization terms that promote ``sparsity'' of spatial model-difference gradients may be expected to improve the recovery of valuable production effects while suppressing undesirable oscillatory artifacts. In this work we present the results of using total-variation (TV) regularization of the model difference in the simultaneous time-lapse FWI and demonstrate the significant improvement of the inversion results compared to alternative methods.

\section{The Method}
Full-waveform inversion is defined as solving the following optimization problem \cite[]{Tarantola,VirieuxOperto}
\begin{equation}
\|\mathbf{M}\mathbf{u}-\mathbf{d}\|_2\;\rightarrow\; \min
\label{eq:fwi}
\end{equation}
where $\mathbf{M},\mathbf{d}$ are the measurement operator and data, $\mathbf{u}$ is the solution of a forward-modeling problem
\begin{equation}
\mathbf{D}(\mathbf{m})\mathbf{u}\;=\;\phi,
\label{eq:fm}
\end{equation}
where $\mathbf{D}$ is the forward-modeling operator that depends on a model vector $\mathbf{m}$ as a parameter, and $\phi$ is a source. The minimization problem (\ref{eq:fwi}) is solved with respect to either both the model $\mathbf{m}$ and source $\phi$ or just the model. In the frequency-domain formulation of the acoustic waveform inversion, the forward-modeling equation (\ref{eq:fm}) becomes
\begin{equation}
-\omega^2 u-v^2(x^1,\ldots,x^n) \Delta u\;=\;\phi(\omega,x^1,\ldots,x^n)
\label{eq:acoustic}
\end{equation}
where $\omega$ is a temporal frequency, $n$ is the problem dimension, and $v$ is the acoustic wave propagation velocity. Values of the slowness $s=1/v$ at all the points of the modeling domain constitute the model parameter vector $\mathbf{m}$. The direct problem (\ref{eq:acoustic}) can be solved in the frequency domain, or in the time domain followed by a discrete Fourier transform in time \cite[]{VirieuxOperto}. The inverse problem (\ref{eq:fwi}) is typically solved using a multiscale approach, from low to high frequencies, supplying the output of each frequency inversion to the next step. 

FWI applications in time-lapse problems seek to recover induced changes in the subsurface model using multiple data sets from different acquisition vintages. For two surveys sufficiently separated in time, we call such data sets (and the associated models) \emph{baseline} and \emph{monitor}.

Time-lapse FWI can be carried out by separately inverting the baseline and monitor models (\emph{parallel difference}) or inverting them sequentially with, e.g., the baseline supplied as a starting model for the monitor inversion (\emph{sequential difference}). Another alternative is to apply the \emph{double-difference} method, with a baseline model inversion followed by a monitor inversion that solves the following optimization problem
\begin{equation}
\|\left(\mathbf{M}^s_m\mathbf{u}_m - \mathbf{M}^s_b \mathbf{u}_b\right) -\left( \mathbf{M}_m\mathbf{d}_m- \mathbf{M}_b\mathbf{d}_b \right)\|_2\;\rightarrow\; \min
\label{eq:dd}
\end{equation}
by changing the monitor model \cite[]{Watanabe,DenliHuang,York,Amir,Raknes}. The subscripts in equation (\ref{eq:dd}) denote the baseline and monitor surveys, $\mathbf{d}$ denotes the field data, and the $\mathbf{M}$'s are measurement operators that project the synthetic and field data onto a common grid. The superscript $s$ indicates the measurement operators applied to the synthetic data.

In all of these techniques, optimization is carried out with respect to one model at a time, albeit of different vintages at different stages of the inversion. In our method we invert for the baseline and monitor models \emph{simultaneously} by solving either one of the following two optimization problems:
\begin{eqnarray}
\label{eq:t1}
\alpha \|\mathbf{M}_b\mathbf{u}_b-\mathbf{d}_b\|_2^2+
\beta \|\mathbf{M}_m\mathbf{u}_m-\mathbf{d}_m\|_2^2+ \\
\label{eq:t2}
\gamma \|\left(\mathbf{M}^s_m\mathbf{u}_m - \mathbf{M}^s_b \mathbf{u}_b\right) -\left( \mathbf{M}_m\mathbf{d}_m- \mathbf{M}_b\mathbf{d}_b \right)\|_2^2+ \\
\label{eq:t3}
\alpha_1 \|\mathbf{W}_b\mathbf{R}_b(\mathbf{m}_b-\mathbf{m}^{\mathrm{PRIOR}}_b)\|_2^2+ \\
\label{eq:t4}
\beta_1 \|\mathbf{W}_m\mathbf{R}_m(\mathbf{m}_m-\mathbf{m}^{\mathrm{PRIOR}}_m)\|_2^2+ \\
\delta \|\mathbf{W}\mathbf{R}(\mathbf{m}_m-\mathbf{m}_b-\Delta \mathbf{m}^{\mathrm{PRIOR}})\|_2^2 \;\rightarrow\; \min,
\label{eq:t5}
\end{eqnarray}
or
\begin{eqnarray}
\label{eq:tl1}
\alpha \|\mathbf{M}_b\mathbf{u}_b-\mathbf{d}_b\|_2^2+
\beta \|\mathbf{M}_m\mathbf{u}_m-\mathbf{d}_m\|_2^2+ \\
\label{eq:tl2}
\gamma \|\left(\mathbf{M}^s_m\mathbf{u}_m - \mathbf{M}^s_b \mathbf{u}_b\right) -\left( \mathbf{M}_m\mathbf{d}_m- \mathbf{M}_b\mathbf{d}_b \right)\|_2^2+ \\
\label{eq:tl3}
\alpha_1 \|\mathbf{W}_b\mathbf{R}_b(\mathbf{m}_b-\mathbf{m}^{\mathrm{PRIOR}}_b)\|_1+ \\
\label{eq:tl4}
\beta_1 \|\mathbf{W}_m\mathbf{R}_m(\mathbf{m}_m-\mathbf{m}^{\mathrm{PRIOR}}_m)\|_1+ \\
\delta \|\mathbf{W}\mathbf{R}(\mathbf{m}_m-\mathbf{m}_b-\Delta \mathbf{m}^{\mathrm{PRIOR}})\|_1 \;\rightarrow\; \min,
\label{eq:tl5}
\end{eqnarray}
with respect to both the baseline and monitor models $\mathbf{m}_b$ and $\mathbf{m}_m$. Problem (\ref{eq:t1}-\ref{eq:t5}) describes time-lapse FWI with $L_2$ regularization of the individual models (\ref{eq:t3},\ref{eq:t4}) and model difference (\ref{eq:t5}) \cite[]{MusaSEP152}. In this work we study the second formulation (\ref{eq:tl1}-\ref{eq:tl5}) that involves $L_1$-regularization of the individual models and their difference.  The terms (\ref{eq:tl1}) correspond to separate baseline and monitor inversions, the term (\ref{eq:tl2}) is the optional double difference term, the terms (\ref{eq:tl3}) and (\ref{eq:tl4}) are optional separate baseline and monitor inversion regularization terms \cite[]{Aster}, and the term (\ref{eq:tl5}) represents regularization of the model difference. In (\ref{eq:tl3})-(\ref{eq:tl5}), $\mathbf{R}$ and $\mathbf{W}$ denote regularization and weighting operators respectively, with the subscript denoting the survey vintage where applicable. If $\mathbf{R}$ is the \emph{gradient magnitude} operator
\begin{equation}
        \mathbf{R}f(x,y,z)\;=\;\sqrt{f^2_x+f^2_y+f^2_z},
        \label{eq:Rgrad}
\end{equation}
then (\ref{eq:tl3}-\ref{eq:tl5}) become \emph{total-variation (TV) seminorms}. The latter case is of particular interest in this work as the minimization of the $L_1$ norm of gradient may promote ``blockiness'' of the model-difference, potentially reducing oscillatory artifacts \cite[]{Aster}.

A joint inversion approach has been applied earlier to the linearized waveform inversion \cite[]{Gboyega}. In \cite{MusaSEP150,MusaSEP152,MusaAGUSEG14}, a simultaneous full-waveform inversion problem (\ref{eq:t1},\ref{eq:t5}) was studied with a single model difference $L_2$ regularization term (\ref{eq:t5}).

An implementation of the proposed simultaneous inversion algorithm requires solving a nonlinear optimization problem with twice the data and model dimensions of problems (\ref{eq:fwi}) and (\ref{eq:dd}). The model difference regularization weights $\mathbf{W}$ and, optionally, the prior $\Delta\mathbf{m}^{\mathrm{PRIOR}}$ may be obtained from prior geomechanical information. For example, a rough estimate of produc\-tion-induced velocity changes can be obtained from time shifts \cite[]{HatchellBourne,BarkvedKristiansen} and used to map subsurface regions of expected produc\-tion-induced perturbation, and optionally provide a difference prior. However, successfully solving the $L_1$-regularized problem (\ref{eq:tl1}-\ref{eq:tl2}) is less sensitive to choice of the weighting operator $\mathbf{W}$. For example, we show below that the TV-regularization using (\ref{eq:Rgrad}) with $\mathbf{W}=1$ recovers non-oscillatory components of the model difference, while the $L_2$ approach would result in either smoothing or uniform reduction of the model difference.

\plot{starting}{width=\columnwidth}
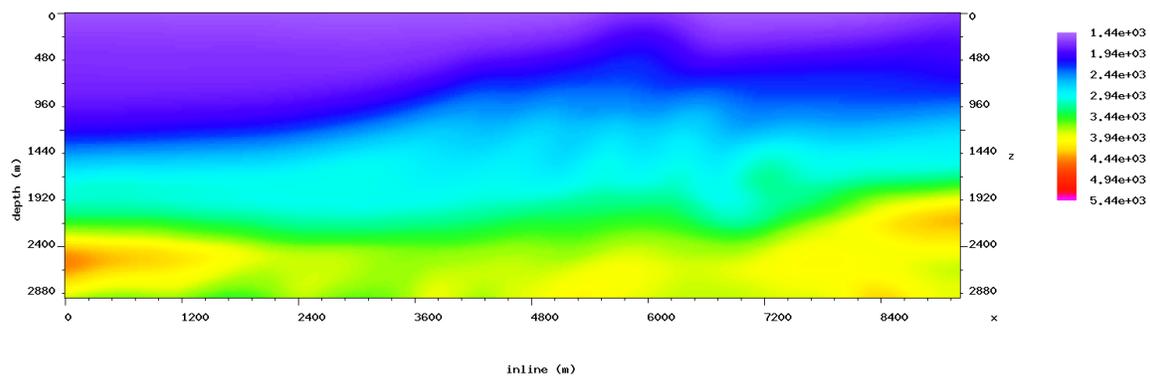
{Starting model used in the inversion.}

In addition to the fully simultaneous inversion, \cite{MusaSEP150,MusaSEP152} proposed and tested a \emph{cross-updating} technique that offers a simple but remarkably effective approximation to minimizing the objective function (\ref{eq:t1}),(\ref{eq:t5}), while obviating the difference regularization and weighting operators $\mathbf{R}$ and $\mathbf{W}$ for problem (\ref{eq:t1},\ref{eq:t5}). This technique consists of one standard run of the sequential difference algorithm, followed by a second run with the inverted monitor model supplied as the starting model for the second baseline inversion
\begin{equation}
        \begin{aligned}
                \mathbf{m}_{\mathrm{INIT}} \rightarrow & \textrm{baseline}\;\textrm{inversion}\rightarrow \textrm{monitor}\;\mathrm{inversion}\rightarrow\\
                                                       & \textrm{baseline}\;\textrm{inversion}\rightarrow \textrm{monitor}\;\mathrm{inversion},
        \end{aligned}
\label{eq:x}
\end{equation}
and computing the difference of the latest inverted monitor and baseline models. Process (\ref{eq:x}) can be considered as an approximation to minimizing (\ref{eq:t1}) and (\ref{eq:t5}) because non-repeatable footprints of both inversions are propagated to both models, canceling out in the difference. Both the simultaneous inversion and cross-updating minimize the model difference by tackling model artifacts that are in the null space of the Fr\'{e}chet derivative of the forward modeling operator. The joint inversion minimizes the effect of such artifacts on the model difference by either minimizing the model difference term (\ref{eq:t5}) in the simultaneous inversion, or  by propagating these artifacts to both models in cross-updating (\ref{eq:x}).  Note that this process is not guaranteed to improve the results of the baseline and monitor model inversions but was only proposed for improving the model difference. \cite{MusaSEP152,MusaAGUSEG14} demonstrated a significant improvement of model difference recovery by both the $L_2$-regularized target-oriented simultaneous inversion and cross-updating compared to the parallel, sequential and double difference techniques. The simultaneous inversion and cross-updating yielded qualitatively similar results within the inversion target.

Here we compare joint simultaneous inversion with a TV-regularized model difference (\ref{eq:tl1},\ref{eq:tl5},\ref{eq:Rgrad}) to parallel difference and cross-updating.

\plot{bb155Hz}{width=\columnwidth}
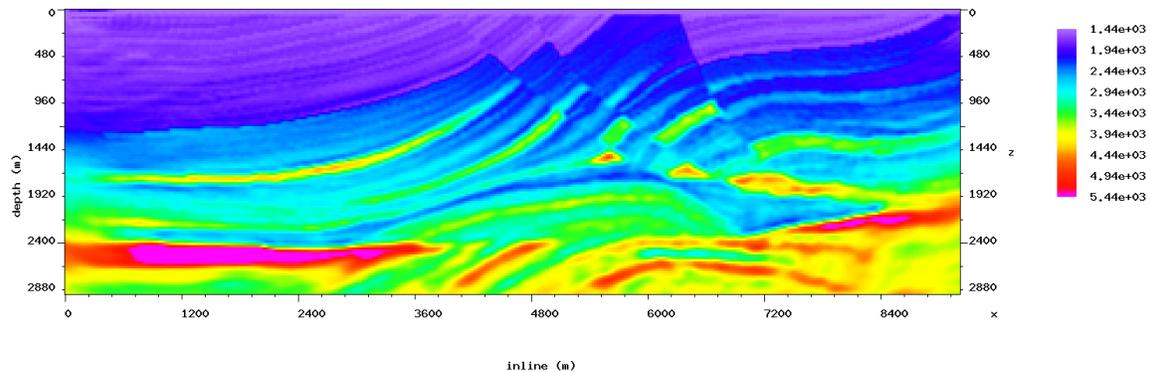
{Baseline model inverted from noise-free synthetic data.}

\section{Numerical examples}

The Marmousi velocity model is used as a baseline, over a 384$\times$122 grid with a 24 m grid spacing. Produc\-tion-indu\-ced velocity changes are modeled as a negative $-150$ m/s perturbation at about 4.5 km inline 800 m depth, and a positive $200$ m/s perturbation at 6.5 km inline, 1 km depth. Additionally, a smoothly varying negative velocity change, peaking at $-50$ m/s, was included above the positive anomaly as shown in Figure~\ref{fig:truediff}.  The whole Marmousi model is inverted, however, only model differences for the section between the approximate inline coordinates 4 km and 6.7 km to the depth of approximately 1.4 km are shown here. The inversion is carried out in the frequency domain for 3.0, 3.6, 4.3, 5.1, 6.2, 7.5, 9.0, 10.8, 12.8, and 15.5 Hz, where the frequencies are chosen based on the estimated offset to depth range of the data \cite[]{SirguePratt}. The baseline acquisition has 192 shots at a depth of 16 m with a 48 m spacing, and 381 receivers at a depth of 15 m with a 24 m spacing. The minimum offset is 48 m. The source function is a Ricker wavelet centered at 10.1 Hz. Absorbing boundary conditions are applied along the entire model boundary, including the surface (thus suppressing multiples). A smoothed true model (Figure~\ref{fig:starting}) is used as a starting model for the initial baseline inversion (and for the initial monitor inversion in the parallel difference). The smoothing is performed using a triangular filter with a 20-sample half-window in both vertical and horizontal directions. The result of inverting the baseline model from the clean synthetic data is shown in Figure~\ref{fig:bb155Hz}. Random Gaussian noise is added to the noise-free synthetic data to produce a noisy data set with 7 dB signal-to-noise ratio. The noisy monitor data set is generated for the model perturbation of Figure~\ref{fig:truediff}, using the same acquisition geometry and source wavelet. The results of baseline model inversion from the clean and 7 dB SNR synthetic data are shown in Figure~\ref{fig:cleanbase} and Figure~\ref{fig:4base}, respectively. Results of model difference inversion from the clean and 7 dB SNR synthetic data sets using various methods are shown in Figures~\ref{fig:clpd},\ref{fig:clx},\ref{fig:cltv} and Figures~\ref{fig:npd},\ref{fig:nx},\ref{fig:ntv}, respectively. The simultaneous inversion objective function contains only terms (\ref{eq:tl1}) and (\ref{eq:tl5}) with no difference prior, i.e., $\Delta\mathbf{m}^{\mathrm{PRIOR}}=0$. The model-difference regularization weights $\mathbf{W}$ in (\ref{eq:tl5}) are set to 1 \emph{everywhere} in the modeling domain. The two terms in (\ref{eq:tl1}) are of the same magnitude and therefore $\alpha$ and $\beta$ are set to 1. Parameter $\delta$ is set to $10^{-5}$ but can be varied for different acquisition source and geometry parameters. The result of the initial baseline inversion is supplied as a starting model for both $\mathbf{m}_b$ and $\mathbf{m}_m$ in the simultaneous inversion. In all the inversions, up to 10 iterations of the nonlinear conjugate gradients algorithm \cite[]{Nocedal} are performed for each frequency. Neither regularization nor model priors are used in single-model inversions (i.e., in the cross-updating and parallel difference methods). \cite{MusaSEP152} demonstrated significant improvement by cross-updating compared to sequential differencing, and rough qualitative equivalence of cross-updating and the $L_2$-regularized simultaneous inversion. Therefore, in this work we compare the TV-regularized simultaneous inversion only against parallel differencing and cross-updating.

\multiplot{2}{cleanbase,4base}{width=.47\columnwidth}
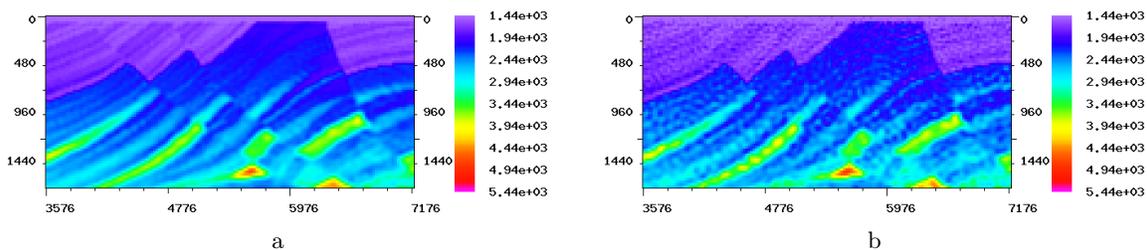
{(a) Target area of the baseline model inverted from clean synthetic. (b) Target area of the baseline model inverted from a 7 dB SNR synthetic. In both cases the baseline model is reconstructed reasonably well, however, errors due to noise are comparable in magnitude to production-induced effects. }

The results of applying cross-updating to the two data sets are shown in Figures~\ref{fig:clx} and \ref{fig:nx}, respectively. The corresponding TV-regularized simultaneous inversion results are shown in Figures~\ref{fig:cltv} and \ref{fig:ntv}.  Since problem (\ref{eq:fwi}) is nonlinear, supplying the result of the highest frequency inversion back to the lowest frequency and repeating the whole inversion cycle for all frequencies may result in achieving a better data fit. In repeated cycles, lower-frequency inversions usually terminate earlier but higher frequencies still deliver model updates. For an objective comparison of the joint inversion with the parallel difference method, the effects of insufficient iteration count are reduced by performing an extra cycle of baseline and monitor inversion (we call this approach ``iterated'' parallel difference \cite[]{MusaSEP152}). The results of applying the iterated parallel difference to the twodata sets are shown in Figures~\ref{fig:clpd} and \ref{fig:npd}. While cross-updating demonstrates certain robustness with regard to uncorrelated noise in the data and computational artifacts (note the significant quantitative improvement of reconstructed difference magnitudes in Figures~\ref{fig:clx} and \ref{fig:nx}), the TV-regularized achieves further signifcant improvement by reducing oscilatory artifacts and honoring both smooth and blocky components of the model difference. 

\plot{truediff}{width=.9\columnwidth}
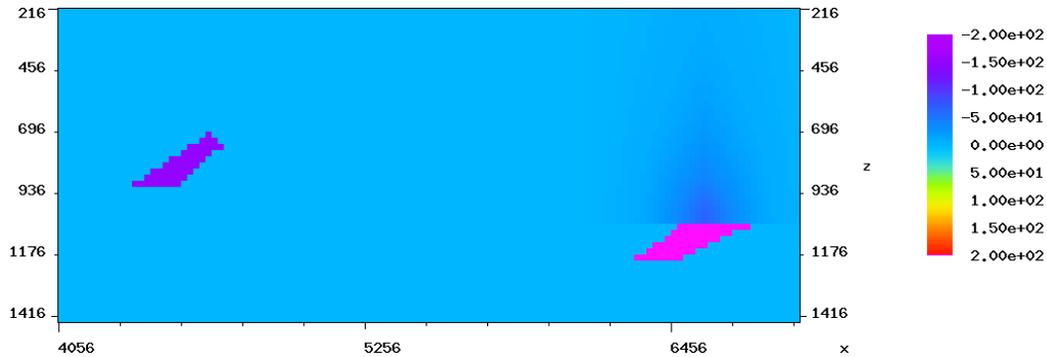
{True velocity difference consists of a negative ($-150$ m/s) perturbation at about 4.5 km inline 800 m depth, and a positive ($200$ m/s) perturbation at 6.5 km inline, 1 km depth.}

\section{Conclusions}

Our new TV-regularized simultaneous inversion technique is a more robust further development of our previous joint inversion method \cite[]{MusaSEP150,MusaSEP152,MusaAGUSEG14}. Use of TV regularization in the simultaneous inversion allows recovery of production-induced changes without specifying variable weighting operator $\mathbf{W}$, and penalizes unwanted model oscillations that may mask useful production-induced changes.

One potentially beneficial extension of this work is using TV regularization of individual models in (\ref{eq:t3},\ref{eq:t5}). This approach may used for both time-lapse and standard FWI, and will be the subject of our next work.

\section{Acknowledgments}

The authors would like to thank Stewart Levin for a number of useful discussions, and the Stanford Center for Computational Earth and Environmental Science for providing computing resources.

\multiplot{3}{clpd,clx,cltv}{width=\columnwidth}
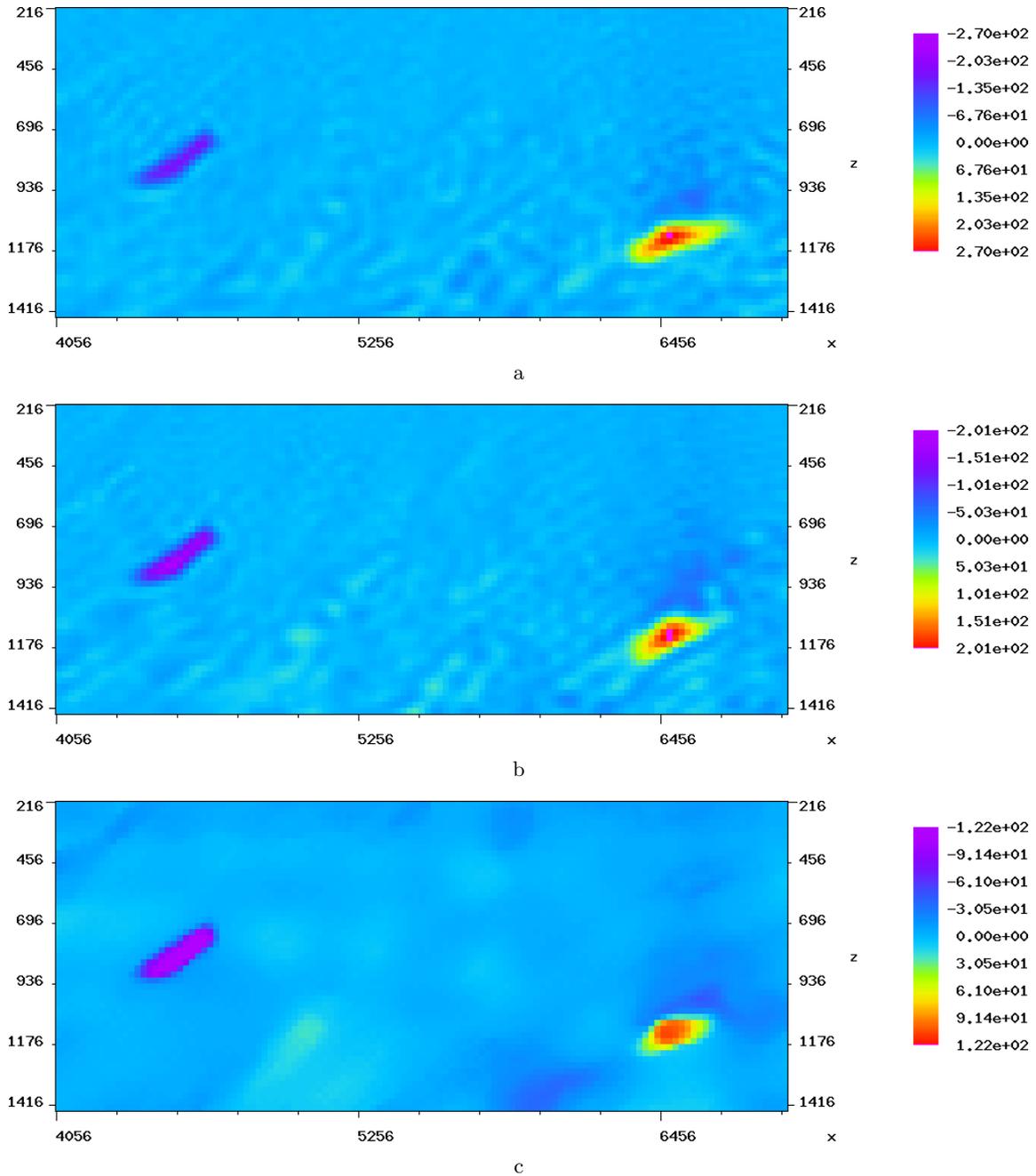
{Model difference inverted from a clean synthetic with matching baseline and monitor acquisition geometries using (a) iterated parallel difference; (b) cross-updating; (c) TV-regularized simultaneous inversion.}

\multiplot{3}{npd,nx,ntv}{width=\columnwidth}
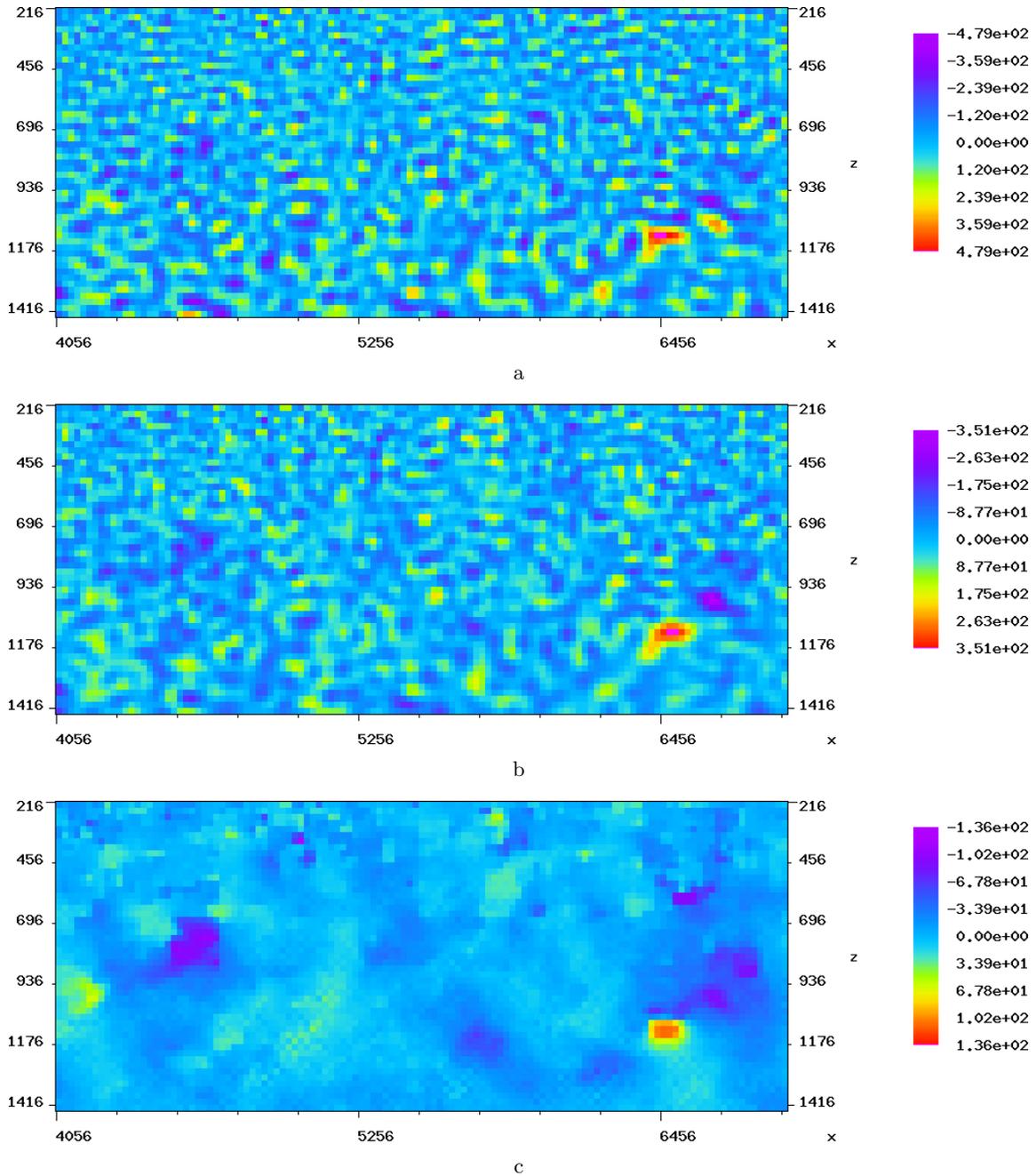
{Model difference inverted from a 7 dB SNR synthetic with matching baseline and monitor acquisition geometries using (a) iterated parallel difference; (b)  cross-updating; (c) regularized simultaneous inversion.}

\bibliographystyle{seg}  
\bibliography{mmbbjointinv}

\end{document}